%% file: main.tex
    \documentclass[11pt,twoside,notitlepage]{article} 
    
	\input{mystyle.sty} 

\begin{document}
\pagenumbering{gobble}

\sloppy 

	
	 	\null\vspace{3cm}
	 	
	\begin{center}
		{\huge \textbf{Efficient Collective Action for Tackling Time-Critical Cybersecurity Threats}\par}
	\end{center}

 		\begin{center}
 			S\'{e}bastien Gillard$^{1,2}$ $^\textrm{[0000-0002-3237-8599]}$, Dimitri Percia David$^{1,3,4}$ $^\textrm{[0000-0002-9393-1490]}$\\ Alain Mermoud$^{3}$ $^\textrm{[0000-0001-6471-772X]}$ and Thomas Maillart$^{1,5}$ $^\textrm{[0000-0002-5747-9927]}$
  				\vspace{0.5cm}
         \end{center}
        
  			\noindent
  			\scriptsize $^{1}$ Information Science Institute, Geneva School of Economics and Management, University of Geneva\\
 			\scriptsize $^{2}$ Department of Defense Economics, Military Academy at ETH Zurich\\
 			\noindent
 			\scriptsize $^{3}$ Cyber-Defence Campus, armasuisse Science and Technology\\ 			\noindent
 			\scriptsize $^{4}$ Institute of Entrepreneurship \& Management, School of Management, University of Applied Sciences of Western\\ \hspace*{2.5mm}Switzerland (HES-SO Valais-Wallis)\\ 
 			\scriptsize $^{5}$ Citizen Cyber Lab, University of Geneva\\
 			\vspace{0.1cm}
    
\begin{center}
    {\bf Presented at the 21st Workshop on the Economics of Information Security (WEIS),\\ 2022, Tulsa, USA\\  \url{https://weis2022.econinfosec.org/}}
\end{center}

\input{abstract}

	\newpage

		\thispagestyle{plain} 
		\newpage 
		\pagenumbering{arabic}
      	\normalsize		
    		    \newpage 
    
	\normalsize
\input{intro}	
\input{background}
\input{data}
\input{theory}
\input{hypotheses}

\input{method}
\input{results}

\input{discussion} 
\input{conclusion}
\input{acknowledgments}

\printbibliography[heading=subbibintoc]

\input{appendices}
\end{document}

%% file: abstract.tex
\begin{abstract}
		\normalsize
        \noindent
        The latency reduction between the discovery of vulnerabilities, the build-up and dissemination of cyber-attacks has put significant pressure on cybersecurity professionals. For that, security researchers have increasingly resorted to collective action in order to reduce the time needed to  characterize and tame outstanding threats. Here, we investigate how joining and contributions dynamics on MISP, an open source threat intelligence sharing platform, influence the time needed to collectively complete threat descriptions. We find that  performance, defined as the capacity to characterize quickly a threat event, is influenced by (i) its own complexity (negatively), by (ii) collective action (positively), and by (iii) learning, information integration and modularity (positively). Our results inform on how collective action can be organized at scale and in a modular way to overcome a large number of time-critical tasks, such as cybersecurity threats.\\

        	   \noindent \keywords{cybersecurity, information sharing, collective action, information integration, economies of scales, Malware Information Sharing Platform (MISP)}
	\end{abstract}

%% file: intro.tex
\section{Introduction}
\label{sec1}

From Computer Emergency Readiness Teams (CERT) established in the nineties \cite{sridhar_cybersecurity_2021}, to information-sharing analysis centers (ISACs) \cite{gal-or_economic_2005}, to bug bounty programs \cite{maillart_given_2017, sridhar_hacking_2021}, collective action has long been used and recognized as key for the gathering, the integration and the sharing of critical cybersecurity information \cite{bohme_back_2016,laube_strategic_2017}. The reason for resorting to information-sharing as a form of collective action stems from the complexity associated with the continuous and somewhat decentralized (e.g., open source software) adaptation of hardware and software in information systems \cite{brady_murphys_1999, stojkovski_whats_2021}. Although the Internet has largely developed through an open source spirit \cite{levy_hackers_2010, benkler_penguin_2011, benkler_wealth_2006} with  significant positive externalities \cite{katz_network_2021, shapiro_information_1999}, information-sharing has remained difficult when it comes to cybersecurity \cite{laube_strategic_2017}. The expansion of threats in volume, severity and span has further challenged information infrastructures. Hence, it has forced further cooperation through information-sharing  \cite{wagner_misp_2016}. While their utility has been somewhat confirmed by their wide adoption, there is a dearth of knowledge regarding how these collective action platforms concretely bring performance when addressing cybersecurity threats. For instance, cybersecurity has become increasingly time-critical and demands ever faster reaction time. Determining the chances that a threat will be fully characterized on time for security officers to act upon before attacks actually start has become crucial \cite{zibak_cyber_2019}.

Here, we investigate $39,639$ threat events contributed by 485 organizations to a MISP information-sharing platform \cite{wagner_misp_2016} operated by the Computer Incident Response Center Luxembourg (CIRCL). We specifically study how collective action unravels through information integration and how it brings significant economies of scale in terms of time needed to fully characterize cybersecurity threats (i.e., performance). We resort to a multivariate cross-sectional regression with ordinary least squares method, and we find that (i) the number of organizations engaged in information-sharing, (ii) their acquired experience in the events completion, (iii) the proportion of information integration and (iv) modularity increase performance.

The remainder of this article is organized as follows. Section \ref{sec:background} covers background from the perspectives of social dilemma, productivity and information integration in collective action in general and for cybersecurity. Section \ref{sec:data} introduces MISP and presents the data. In Section \ref{sec:theory}, we introduce the theoretical framework followed by research hypotheses in Section \ref{sec:hypotheses}. Section \ref{sec:method} describes the methodological approach. Results are presented in Section \ref{sec:results} and discussed in Section \ref{sec:discussion} before concluding in Section \ref{sec:conclusion}.

%% file: background.tex
\section{Background}
\label{sec:background}
Knowledge sharing in cybersecurity has been considered as a crucial way to overcome number of vulnerabilities \cite{mell_common_2006} and threats \cite{sridhar_cybersecurity_2021}. It is however bound to limiting factors on the one hand, such as social dilemma, as well as enhancing return-on-scale effects on the other hand. Here, we review the literature on (i) social dilemma and productivity of collective action, and on (ii) challenges associated with information integration. We then review the state-of-the-art research in (iii) information sharing for cybersecurity. 
    
\subsection{Social Dilemma and Productivity in Collective Action}
\label{social_dilemma}

According to Olson's logic of collective action, small communities are more able to provide collective goods \cite{olson_logic_1971}. The central argument is that minor interests will be over-represented and diffuse majority interests trumped, due to a free-rider problem \cite{anesi_moral_2008}. This free-riding effect is stronger for larger groups \cite{esteban_collective_2001}. For instance, while Dejean et al. \cite{dejean_olsons_2010} found a positive relation between the size of a community and the amount of collective good provided, they paradoxically also found a decreased propensity by individuals to cooperate as the size of the community increases. Yet, there is overwhelming evidence that large crowds can be organized in order to establish successful online collective action. Examples include peer-to-peer networks \cite{asvanund_empirical_2004, dejean_olsons_2010},  Wikipedia \cite{klein_virtuous_2015},  Stack Overflow \cite{wang_empirical_2013}, communities of open source software developers \cite{hippel_open_2003,sornette2014much}. 
The Dejean et al. paradox may at least partially resolved by considering that (i) the distribution of effort is highly skewed, with few contributors providing most effort, and (ii) the dynamics of contribution are highly non-linear \cite{sornette2014much,maillart_aristotle_2019,muric_collaboration_2019}. Taken together, these phenomena are associated with positive return-on-scale of production \cite{sornette2014much}, which may be hindered by coordination costs \cite{scholtes_aristotle_2016}. Super-linear productivity has been debated at length in organization and management sciences. Investigations of how the number of members, temporal dynamics of events generated can influence positively outputs in way that is greater than the sum of the outputs related to each element of the system (i.e., exhibiting super-linear growth patterns). Research has successfully delivered hints to improve the performance of organization \cite{tziner_effects_1985, sundstrom_work_1990, cohen_what_1997, neuman_team_1999} by fine-tuning complementary mechanisms within the organization \cite{ennen_whole_2010}, which also foster innovation \cite{sacramento_team_2006}.

\subsection{Information Integration and Modularity}
\label{sec:knowledge_integration}

One key aspect of generating return-on-scale in knowledge production is information integration. The management of information resources has become central to organizations \cite{nonaka_dynamic_1994}, so that knowledge appears as an utmost strategic resource \cite{grant_prospering_1996}. For instance, there is growing evidence in science that greater teams create more impacting knowledge \cite{wuchty_increasing_2007}. If knowledge is so important, the fundamental capability of an organization has to be considered as the specialized knowledge of each organization member. Its integration shall provide a competitive advantage \cite{grant_prospering_1996,lawrence_differentiation_1967}. With the emergence of virtual exchanges, firms are increasingly seen as distributed knowledge systems \cite{majchrzak_catalyst_2018}. Yet, new interaction methods present various new constraints in term of mutual understanding, contextual knowledge or techniques (e.g., memory, connectivity), which lead to asymmetries in information integration.

In this respect, the tremendous development of online collaboration platforms, as tools for governance strategy and knowledge management, highlights the importance of information-sharing \cite{safa_information_2016}. These platforms promote knowledge transfer by generating modular collaborative units \cite{mockus_case_2000}. One may consider that individuals, or groups of individuals, composing a subsystem (i) bring added value in their own specific field (differentiation), in order to (ii) produce a complex good by pooling together this added value (integration). Following Arrow \& Debreu \cite{kenneth_j_arrow_and_gerard_debreu_existence_1954}, differentiation and integration have been a focal point in optimizing the structure of organizations \cite{ravasi_organising_2001, huang_knowledge_2003}. In fact, differentiation considers segments of a system into subsystems. Each subsystem develops a part of a task, while the integration focuses on the interactions between these subsystems in order to accomplish the entire task \cite{lawrence_differentiation_1967,engel_integrated_2018}. Recently, Engel and Malone used the theory of consciousness as information integration \cite{tononi_consciousness_1998} to measure information integration computer systems and on collaborative platforms \cite{engel_integrated_2018}.

\subsection{Collective Action and Information Integration for Cybersecurity} 

As early as twenty years ago, the first Computer Emergency Readiness Teams (CERT) and Information Sharing and Analysis Centers (ISACs) have been established as a central resource for sharing information on cybersecurity threats to critical infrastructures \cite{zheng_cyber_2015}. Nowadays, threat intelligence platforms help organizations aggregate, correlate, and analyze threat data from multiple sources in (almost) real-time to support defensive actions \cite{he_perspectives_2018}. Further, open source solutions have been proposed as a counterweight to cyber-criminals successfully working together \cite{bohme_back_2016}.
The swift evolution of cyber-threats has forced organizations and governments to develop new strategies \cite{meier_feedrank_2018} in order to reduce the risks of security breaches \cite{safa_information_2016}. Although information sharing
is an interesting way to enhance cybersecurity, it is believed to be thwarted by social dilemma. Without trust, commitment and shared vision between stakeholders, organizations are reluctant to share information due to the fear of disclosure, reputation risk or loss of competitive power \cite{mermoud_share_2019}. As such, information-sharing can be considered as a marketplace on which transactions occur and knowledge is transferred \cite{perciadavid_knowledge_2020}. However, human beings have a tendency to not optimize organizational goals \cite{mermoud_governance_2019} and – in the case of collective action – might adopt behaviors that are not conducive to the overall goal of sharing information \cite{laube_strategic_2017}. As a consequence, cybersecurity professionals share probably less information than desirable, leading to a knowledge asymmetry to the advantage of the attackers \cite{laube_strategic_2017}. In particular, stakeholders strategically select their contributions to share (i.e., quantity and quality), leading to truncated and imperfect information sharing. Yet, specially crafted forms of cybersecurity information-sharing platforms have developed, such as bug bounty marketplaces. These platforms act as a trusted third-party between security researchers and software editors \cite{maillart_given_2017}. Further, in cybersecurity, resource belief, usefulness belief, and reciprocity belief are all positively associated with knowledge absorption, whereas reward belief is not \cite{perciadavid_knowledge_2020}. These empirical results show that functional cybersecurity information-sharing indeed requires to overcome social dilemma and goes beyond simple reward expectations, but foremost requires that information-sharing is efficient in a context that increasingly requires to address time-critical threats.

%% file: data.tex
\section{Data}
\label{sec:data}

To understand the nuts and bolts of cybersecurity information-sharing, we resort to {\it MISP Project},\footnote{\url{https://www.misp-project.org/}} a popular open source platform, which is used e.g., by the North Atlantic Treaty Organization (NATO).\footnote{\url{https://misp.ncirc.nato.int}} MISP stands for {\it Malware Information Sharing Platform and Threat Sharing}. Although it carries the word malware in its name, MISP is a threat intelligence platform on which people can share, store and collaborate on all sorts of incidents (e.g., COVID-19 MISP community,\footnote{\url{https://covid-19.iglocska.eu}} but primarily cybersecurity threats. These threats (i.e., events) are characterized by indicators of compromise (i.e., attributes), which are contributed by a multitude of organizations.

There are advantages in using MISP as an object of research. First, it is an open source software. This allows to understand in much detail how the platform is designed and works. Second, a number of threat information sharing communities use MISP to share relatively openly their threat intelligence. Here, we use the whole history of a MISP instance maintained by the Computer Incident Response Center Luxembourg (MISP CIRCL), i.e., the Luxembourg CERT.

\begin{figure}[ht]
    \centering
    \includegraphics[width=\textwidth]{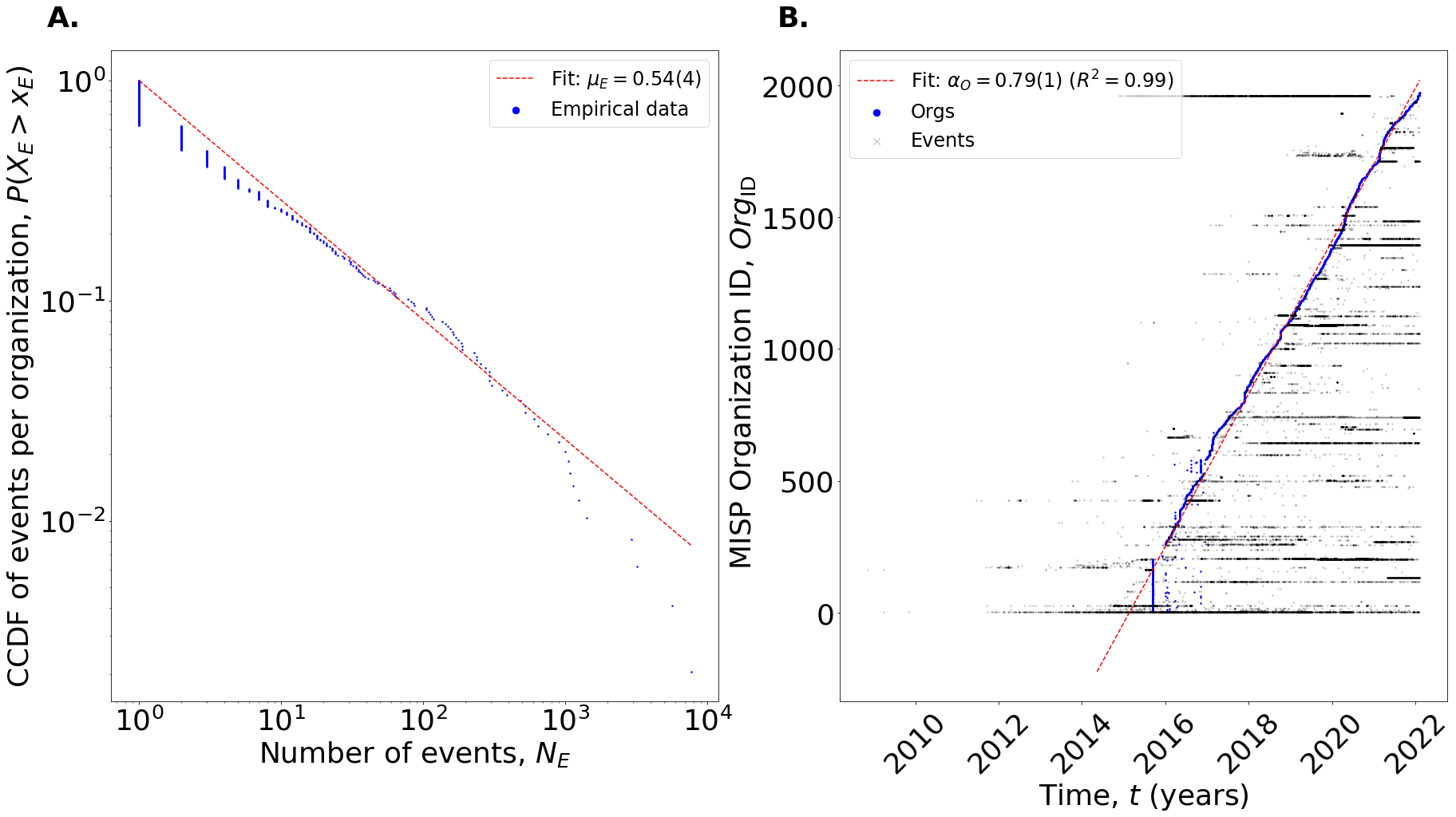}
    \caption{{\bf A.} Complementary cumulative distribution function (CCDF) of events per contributing organization, which is best described by a power law distribution $P(X_E>x_E) \sim 1/{x_E}^{\mu_E}$ with $\mu_E = 0.54(4)$. The fit and the goodness-of-fit, provided by the Kolmogorov-Smirnov statistics test, are obtained with the \texttt{Python} library \texttt{plfit}. {\bf B.} Curve of the joining organizations (in blue) has followed, after the September 14, 2015, the presumed date of official start, a linear growth with slope $\alpha_O = 0.79(1)$, ($R^2 = 0.99,~p\textrm{-value} < 10^{-2}$). The events contributed by the organizations have been added (in dark gray), the distribution shows the heterogeneity of organizations efforts.} 
    \label{fig:org}
    \end{figure}

\begin{figure}[ht]
    \centering
    \includegraphics[width=\textwidth]{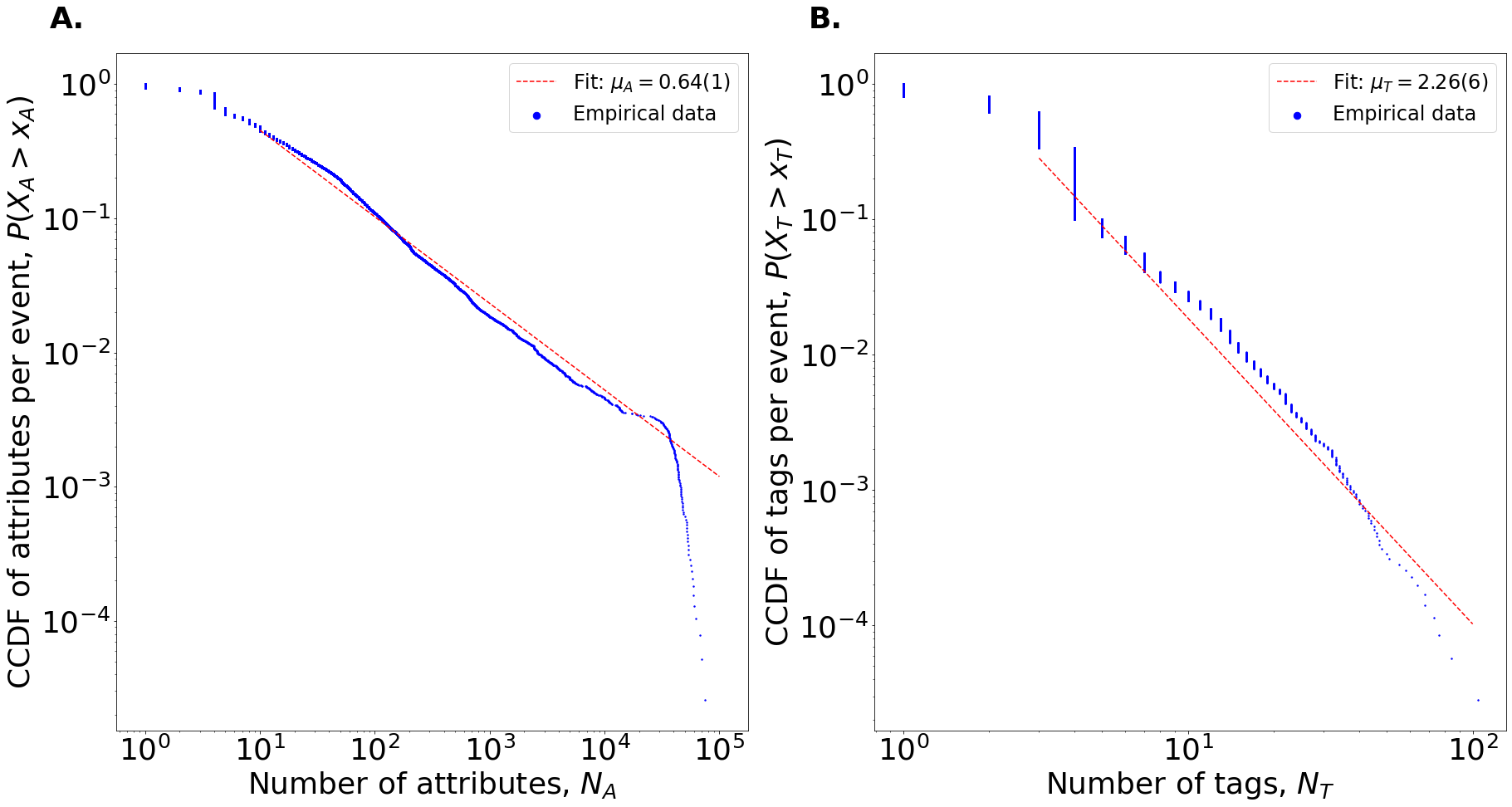}
    \caption{{\bf A.} Complementary cumulative distribution function (CCDF) of attributes encapsulated in an event, which is best described by a power law distribution $P(X_A>x_A) \sim 1/{x_A}^{\mu_A}$ with $\mu_A = 0.64(1)$. {\bf B.} CCDF of tags attached to an event which is best described by a power law distribution $P(X_T > x_T) \sim 1/x_T^{\mu_T}$ with $\mu_T = 2.26(6)$. The fits and the goodness-of-fits, provided by the Kolmogorov-Smirnov statistics test, of panels A and B are obtained with the \texttt{Python} library \texttt{plfit}.}
    \label{fig:CCDFs}
    \end{figure}

As of February 8, 2022, the MISP CIRCL instance is a community of $1,908$ organizations (respectively $4,013$ users), which have contributed $39,639$ events, $9,099,685$ attributes and $3,786$ tags since November 10, 2008. Table \ref{tab:top_10_contributors} shows the ten most involved organizations. One can see that the number of events contributed by organizations is highly skewed. Indeed, Figure \ref{fig:org}A shows that the complementary cumulative distribution function exhibits a power law $P(X_{E}>x_{E}) \sim 1/x_{E}^{\mu_{E}}$ with $\mu_{e} = 0.54(4)$ (c.f., Appendix \ref{sec:ExploA} for details on the fitting method). One may additionally note that $1,423$, i.e., around $75\%$, of organizations do not participate in sharing threat information as a collective good with the broad MISP CIRCL community. These organizations may however consume information or share threat information privately within informal sub-groups, which cannot be observed. Similarly to $P(X_E > x_{E})$, the distributions of attributes $P(X_{A} > x_{A})$ and tags $P(X_{T} > x_{T})$ per event, depicted in Figure \ref{fig:CCDFs}, follow power laws with exponents respectively $\mu_A = 0.64(1)$ (with an upper cut-off around $A_{upper} = 10^5$) and $\mu_{T} = 2.26(6)$. It is additionally important to consider that only $22,423$ (i.e., around $57\%$) events have been marked as completed, suggesting that either threat analysis is complicated or that users tend to forget to formally close a large number of events. The cumulative number of tags $N_{T,\textrm{cum}} = 116,407$ used is bigger than the unique tags amount $N_{T_U} = 3,786$. Thus, there is a massive reuse of already existing tags.
    
\begin{table}[ht]
    \centering
\begin{tabular}{|l|l|l|r|r|}
\hline
rank & org ID & \# users        & \multicolumn{1}{l|}{\# events contributed} & \multicolumn{1}{l|}{percentage of total events} \\ \hline
1    & 1092   & 8          & 7,682                                     & 19.38\%                                         \\
2    & 1395   & 2              & 5,637                                     & 14.22\%                                         \\
3    & 1960   & 3          & 3,214                                     & 8.11\%                                          \\
4    & 2      & 31             & 2,939                                     & 7.41\%                                          \\
5    & 1857   & 3          & 1,411                                     & 3.56\%                                          \\
6    & 201    & 8              & 1,247                                     & 3.15\%                                          \\
7    & 1713   & 1              & 1,141                                     & 2.88\%                                          \\
8    & 698    & 2          & 1,077                                     & 2.72\%                                          \\
9    & 204    & 56         & 1,060                                     & 2.67\%                                          \\
10   & 643    & 12             & 998                                       & 2.52\%                                          \\ \hline
     &        & \textbf{Total} & 26,406                                    & 66.62\%                                         \\ \hline
\end{tabular}
    \caption{$10$ of $1,908$ organizations have contributed  $66.62\%$ of the $39,639$ events, bringing further evidence of the heavy-tailed nature of the distribution of contributions by organizations in MISP CIRCL.}
    \label{tab:top_10_contributors}
    \end{table}
    
We further observe that organizations have joined MISP CIRCL following an almost perfect linear relation $N_O(t) \sim \alpha_O \cdot t $ with $\alpha_O = 0.79(1)$ ($R^2 = 0.99$ and $p<10^{-2}$) with $161$ organizations initially joining MISP CIRCL instance on September 14, 2015, the presumed date of official start. Figure \ref{fig:org}B, not only shows the almost linear organization joining rate, but also how many events each organization has contributed over time. One see that the contribution effort is highly heterogeneous. It is also worth noting that event contributions started on November 10, 2008, long before the first organizations joined MISP CIRCL instance. This can be explained in the following way: organizations run first their MISP instance locally. At some point, they join the MISP CIRCL community and share at once all their non-private threat intelligence, yet with the nominal event timestamp, which may well be in the past. Also, it is likely that the linear organization joining function may be the result of a highly vetted joining process, controlled by CIRCL.


\subsection{Reduction of the Completion Time of Events \texorpdfstring{$\Delta t_C$}{Delta t}}

Following the method described in the Appendix \ref{sec:ExploA}, we can treat the data and, from them, generate the Figure \ref{fig:Deltat}B. As explained in the appendix, by playing with the axis, we remark that when the axes are in linear-logarithmic scale, the data depict two straight lines. From this observation, we can deduce that $\Delta t_C(t)$ follows an exponential decrease in phase. By applying a binning by month and computing the mean value $\overline{\Delta t_C}$ for each bin, we see a first phase that extends from 2011 to 2020 which decrease slower than the second phase from 2020 to today. By applying the linear regression on the data, according to the equation $\eqref{eq:lin-log}$, we confirm that $\Delta t_C$ exhibits an exponential decrease:
    \begin{align}
        \Delta t_C(t) =
        \begin{cases}
        \sim 10^{\beta_{\Delta}^1 \cdot t}, \textrm{ for } t \in [2011,2020[, \\
        \sim 10^{\beta_{\Delta}^2 \cdot t}, \textrm{ for } t \in [2020,2022],
        \end{cases}
    \end{align}
    \noindent
    where
    \begin{quote}
        \begin{itemize}
            \item[--] $\beta_{\Delta}^1 = (- 6.32 \pm 0.91) \times 10^{-3}$ is the exponential decrease of the first part regression and
            \item[--] $\beta_{\Delta}^2 = (-7.12 \pm 0.59) \times 10^{2}$ is the exponential decrease of the second part regression.
        \end{itemize}
    \end{quote}

The fit from the linear is of high quality since its Pearson's determination coefficient $R^2 =0.86$ and its $p\textrm{-value}< 10^{-2}$.
Hence, the time $\Delta t_C$ to complete an event decreases over time, indicating an improvement of performances of the MISP CIRCL instance.

\begin{figure}[ht]
\centering
\includegraphics[width=\textwidth]{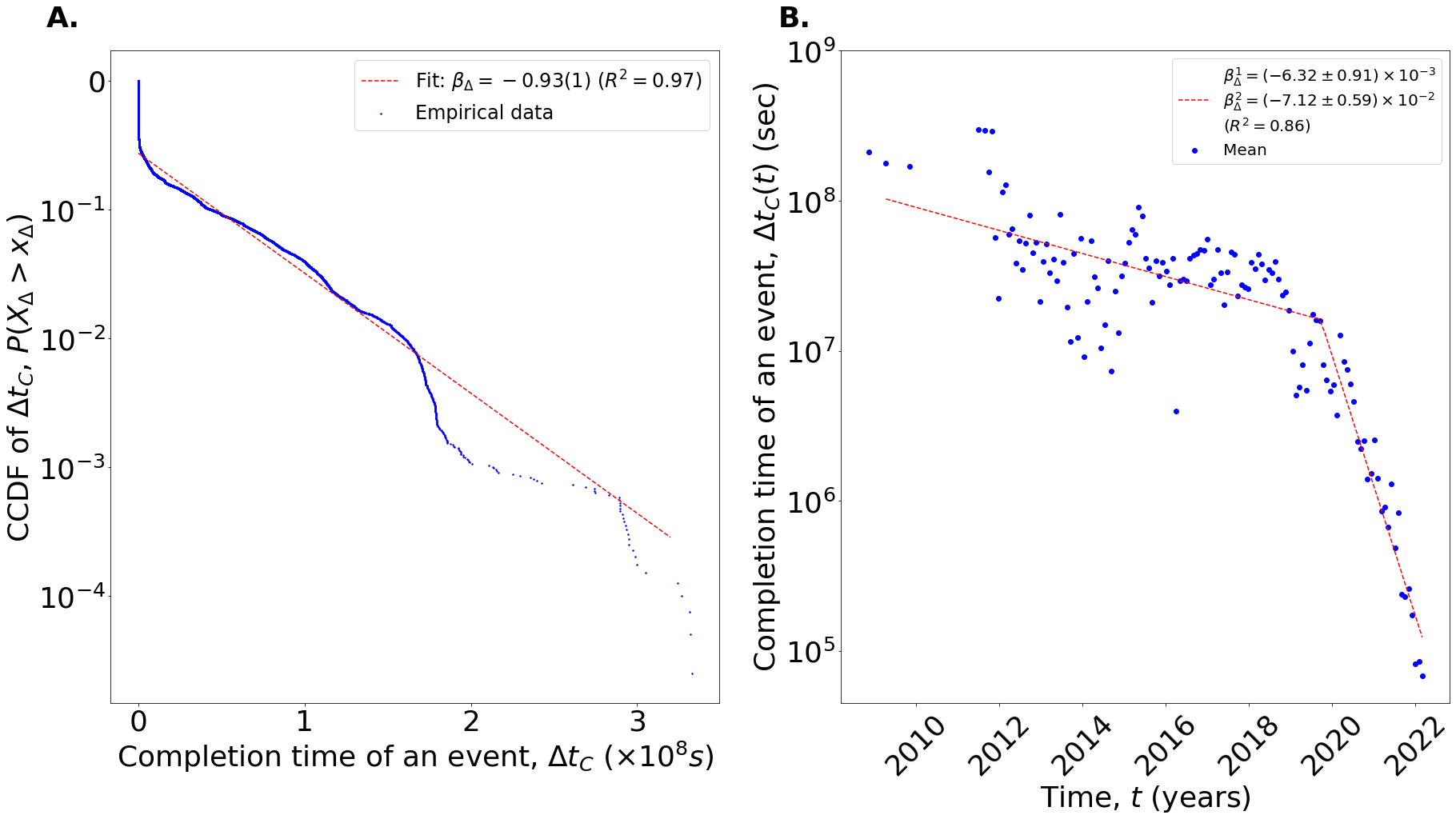}
\caption{{\bf A.} Complementary cumulative distribution function (CCDF) of the completion time $\Delta t_C$, which is best described by a decreasing exponential distribution $P(X_{\Delta} < x_{\Delta}) \sim 10^{\beta_{\Delta}}$ with $\beta_{\Delta} = -0.93(1)$. {\bf B.} Completion time $\Delta t_C$ of events over the time. The data (blue dots) represents the mean value of $\Delta t_C$ binned monthly. The data depict an exponential decrease in two phases, fitted by linear regression (dashed red line), $\Delta t_C(t) \sim (-6.32 \pm 0.91) \times 10^{-2}$ for $t \in [2011,2020[$ and $\Delta t_C(t) \sim (-7.12 \pm 0.59) \times 10^{-2}$ for $t \in [2020,2022]$ ($R^2 = 0.86,~p\textrm{-value} < 10^{-2}$). The fits and their goodness-of-fits, provided by the Pearson's coefficient of determination $R^2$ and the $p$-value for the Wald test, of panels A and B are obtained with the \texttt{Python} library \texttt{scipy.stats.linregress}.}
\label{fig:Deltat}
\end{figure} 

%% file: theory.tex
\section{Theoretical Framework}
\label{sec:theory}
Collective action is thought to be a fundamental tool to overcome sprawling and increasing time-critical cybersecurity threats \cite{mermoud_three_2019,bouwman_governance_2018,wagner_cyber_2019}. Yet, despite numerous studies of online platforms fostering collective action \cite{bouwman_helping_2022,mccoll_performance_2014}, very little evidence has been uncovered linking the organisation of collective action with group performance as an output. By investigating the MISP threat management platform run by the Computer Incident Response Center Luxembourg (CIRCL), we have a unique chance to better understand how collective action is organized to tackle time-critical cybersecurity threats.

We posit that the  performance of collective platforms devoted to the resolution of time-critical tasks at scale, such as MISP, pull from progressively building a knowledge and action environment, made of organizations, which contribute to the resolution of events and, at the same time, bring returns of scale through (i) gaining own experience and (ii) sharing and integrating knowledge, which is associated with increased performance. We further posit that, in order to offset decreasing return-of-scale due to increased groups size and coordination costs \cite{scholtes_aristotle_2016}, the organization of collective action must adapt in a modular way \cite{baldwin_architecture_2006}, as it has already been witnessed in several open source projects \cite{narduzzo_role_2005,langlois_hackers_2008}.

We test our theory of {\it collective action for tackling time-critical tasks}, through a set of three hypotheses and six sub-hypotheses to understand how time completion performance is achieved for events, given (i) the nature of event, (ii) the collective action environment and (iii) the knowledge integration environment at the time of event arrival (c.f., section \ref{sec:hypotheses}). We proceed with an exploratory approach to test our theory by resorting to a multivariate cross-sectional regression with ordinary least squares method (c.f., sections \ref{sec:method} and \ref{sec:results}).

%% file: hypotheses.tex
\section{Hypotheses}
\label{sec:hypotheses}

To explain how event completion time has evolved, we consider their {\it intrinsic nature}, i.e., number of attributes and tags required to characterize events, the {\it overall collective action environment} and how {\it knowledge is integrated}. We hypthesize that these three overall factors significantly influence collective action performance, in terms of improved completion time in characterizing threat events.

\subsection{Event Complexity Hinders Performance (H1)}

First, events are not all equal: while some are fairly simple and require limited input in terms of attributes and of categorization with tags, others are more complex and require more effort. As shown on Figures \ref{fig:CCDFs}A and \ref{fig:CCDFs}B, the distribution of respectively attributes and tags is heavy-tailed: while a majority of events have a limited number of attributes (resp. tags), some carry a large numbers of attributes (resp. tags), presumably affecting the time required to complete the characterization of an event. Hypothesis 1 states:\\ 

\textbf{H1}: \textit{The number of attributes and tags per event negatively influences performance.}\\

\subsection{Collective Action Improves Performance (H2)}
 
We consider how collective action at scale affects positively or negatively performance. Namely, there are conflicting views on whether having more stakeholders (e.g., contributors, organizations) joining collective action is likely to enhance or hinder performance \cite{olson_logic_1971, sornette2014much,scholtes_aristotle_2016, maillart_aristotle_2019,muric_collaboration_2019}. Yet, to exist and be sustainable, collective action necessarily needs to bring economies of scale of some form, which in turn would attract more contributors. Conversely, having more participants should bring marginally increasing performance. Therefore, we aim to test the following hypothesis:\\

\textbf{H2a}: \textit{The overall performance increases with the number of organizations participating in collective action}.\\

Yet, as already shown in \cite{kuypers_designing_2018}, the ongoing collective action workload is likely to affect negatively performance, by increasing completion time. Therefore, our second hypothesis states:\\

\textbf{H2b}: \textit{Given a focal event, the number of simultaneously open events decreases performance.}\\

\subsection{Knowledge Integration Increases Performance (H3)}
 
Having more contributors does not necessarily imply economies of scale \cite{scholtes_aristotle_2016}. Economies of scale may rather be generated by ``the whole is more than the sum of its parts'' mechanisms \cite{sornette2014much}, which may stem from productive integration of information  \cite{oizumi_phenomenology_2014, malone_superminds_2019,engel_integrated_2018} as a single entity \cite{sornette2014much} or through the efficient communication of several modular sub-systems \cite{barrett_practical_2011,baldwin_design_2000}, which in turn may even mitigate free-riding \cite{baldwin_architecture_2006}. Here, we recognize that the first form on knowledge integration occurs through experience as {\it learning} within organizations  \cite{argote_organizational_2011}, and one may expect that an organization having accumulated experience in characterizing a large number of threat events is likely to perform better on new events, therefore :\\

\textbf{H3a}: \textit{More experienced organizations solve events faster.}\\

On MISP instances, collective action goes beyond coordinating time-critical tasks. As people and organizations contribute, a large corpus of knowledge is built as a library of events, attributes, and tags. In turn, by design of MISP software, this information can be easily reused to quickly characterize new events, proposing matching possibilities according to the preliminary entries.

Hence, the reuse of knowledge simplifies the emission of attributes and the knowledge is integrated by the creator of the new events. These new events are thus composed of a certain percentage of {\it inherited} attributes which are likely to impact positively performance:\\ 

\textbf{H3b}: \textit{The reuse of tags and attributes from existing events contributes positively to performance in the completion of new events.}\\

The capacity of an entity to integrate knowledge is tightly related to its modular organization \cite{tononi_consciousness_1998,narduzzo_role_2005,baldwin_architecture_2006}. As MISP clusters of events or attributes, called ``Galaxies'', were progressively introduced and developed on MISP CIRCL, we have an opportunity to test for modularity. We therefore formulate the following hypothesis:\\

\textbf{H3c}: \textit{Modularity in collective action positively influences performance.}\\

By testing these three hypotheses (and six sub-hypotheses), we expect to gain robust insights on how collective action on MISP brings performance in terms of characterizing time-critical cybersecurity threats.

%% file: method.tex
\section{Method}
\label{sec:method}

We proceed to validate our theory through the testing of three hypotheses, divided in six sub-hypotheses (c.f., Section \ref{sec:hypotheses}).  For this, we specify an econometric model with {\it completion time} as the main dependent variable representing the key performance indicator in our posited {\it theory of collective action for tackling time-critical threats} (c.f., Section \ref{sec:theory}).

We define the following set of events,

\begin{align}
    \Omega_e = \{e|e \leq N_e, e \in \mathbb{N}^{\ast}\},
\end{align}
\noindent
where $N_e$ corresponds to $22,423$ events, which have explicitly been marked as completed (i.e., with field $Analysis = 2$, see section \ref{sec:data}). For each event, we define $\Delta t_{C,e}$ the completion time of events as 
\begin{equation}
    \Delta t_{C,e} = t_{f,e} - t_{c,e},
\end{equation} 
\noindent
with $t_{c,e}$ the event creation date and $t_{f,e}$ the last event modification.  

To determine the relation between the dependent variable, i.e. the completion time $\Delta t_{C,e}$ for the events, we proceed to a multivariate cross-sectional regression \cite{asteriou_applied_2015}. Specifically, we investigate if completion time $\Delta t_{C,e}$ for the events can be explained by the selected explanatory variables. The corresponding \texttt{Python} variable is \texttt{CompletionT}. For each event $e$, the multivariate cross-sectional regression writes:

\begin{align}
    \log(\Delta t_{C,e}) &= \zeta + \sum_{k=1}^{N_k} \cdot \sum_{e=1}^{N_e} \kappa_{k} \cdot \log(Z_{k,e}) + \varepsilon_e,
    \label{eq:EcoModel}
\end{align}
\noindent
with:
\begin{quote}
   \begin{itemize}
       \item[--] $\Delta t_{C_e}$ : time completion for event $e$, 
       \item[--] $\zeta$ : constant,
        \item[--] $N_k$ : number of explanatory variables,
       \item[--] $\kappa_{k}$ : autoregressor parameter  corresponding to $Z_{k,e}$,
       \item[--] $Z_{k,e}$ : k-th explanatory variable for event $e$,
       \item[--] $\varepsilon_e$ : error term (i.e., $\log(\Delta t_{C,e}) - \log(\widehat{\Delta t_{C,e}})$). 
   \end{itemize}
\end{quote}

This multivariate cross-sectional regression is performed with the ordinary least squares (OLS) method. The choice of this model is  adapted to deal with data without time series, which is the case here. Then, the explicated and explanatory variables are linked with a set of points in time. This set of points in time is given by the creation $t_{c,e}$ of the different $e$ and contains $22,423$ elements, corresponding to the number of completed elements $N_e$ considered.  
Thanks to this model, it is easy to consider all chosen independent variables. However, due to the heavy-tailed behaviour of the variables and their difference of magnitude (see Section \ref{sec:data}), we transform the variables in logarithm in base of 10 \cite{benoit_linear_2011}. However, the results are indicated as a percentage change of $\Delta t_{C,e}$ when $Z_{k,e}$ varies by a certain percentage \cite{benoit_linear_2011}.

We specify the following explanatory variables in relation with the formulated hypotheses (c.f., Section \ref{sec:hypotheses}). To test hypothesis {\bf H1} (i.e., {\it event complexity hinders performance}), we resort to two explanatory variables:
\begin{quote}
\begin{itemize}
    \item[--] $N_{A,e}$: the number of attributes per event $e$. The corresponding \texttt{Python} variable is \texttt{AttrCount}, which is expected to positively influence \texttt{CompletionT} (i.e., reduce performance).
    \item[--] $N_{T,e}$: the number of tags per event $e$, The corresponding \texttt{Python} variable is \texttt{NTags}, which is expected to positively influence \texttt{CompletionT} (i.e., reduce performance). 
\end{itemize}
\end{quote}

To test hypothesis {\bf H2} (i.e., {\it collective action improves performance}), we resort to two explanatory variables:

\begin{quote}
\begin{itemize}
\item[--]$N_{O,e}$ stands for the number of organizations listed on MISP CIRCL at the creation $t_{c,e}$ of event $e$. The corresponding \texttt{Python} variable is \texttt{CumOrgs}. \texttt{CumOrgs} is expected to negatively influence \texttt{CompletionT} (i.e., increase performance) and to demonstrate the overall benefits of collective action for tackling time criticial threats ({\bf H2a}). 
\item[--] $E_{\textrm{sim},e}$ is the number of simultaneously open events on MISP CIRCL at the creation $t_{c,e}$ of event $e$. The corresponding \texttt{Python} variable is \texttt{SimEvents}, which is expected to positively influence \texttt{CompletionT} (i.e., reduce performance) and to show that collective action performance is bound to circumstantial operational constraints associated with time as a scarce resource ({\bf H2b}) \cite{maillart_quantification_2011,kuypers_designing_2018}.
\end{itemize}
\end{quote}

To test hypothesis {\bf H3} (i.e., {\it knowledge integration increases performance}), we resort to three explanatory variables:

\begin{quote}
\begin{itemize}
    \item[--] $E_{C,e}$ takes into account the number of already completed events by the organizations at the creation $t_{c,e}$ of a new event $e$ on their behalf. The corresponding \texttt{Python} variable is \texttt{CumCompE}, which is expected to negatively influence \texttt{CompletionT} (i.e., increase performance) ({\bf H3a}).
    \item[--] $I_{{\%A},e}$ is the inherited percentage of attributes per event $e$. The corresponding \texttt{Python} variable is \texttt{InhPer}, which is expected to negatively influence \texttt{CompletionT} (i.e., increase performance) ({\bf H3b}).
    \item[--] $N_{G,e}$ counts the number of galaxies created on MISP CIRCL instance at the creation $t_{c,e}$ of the $e$. The corresponding \texttt{Python} variable is \texttt{NbGalaxies}, which is expected to negatively influence \texttt{CompletionT} (i.e., increase performance)  ({\bf H3c}). 
    \item[--] $N_{E_G,e}$ considers the number of events in its corresponding aforementioned galaxy at the creation $t_{c,e}$ of a new event $e$ in this galaxy. The corresponding \texttt{Python} variable is \texttt{NbEventsinhisG}, which is expected to negatively influence \texttt{CompletionT} (i.e., increase performance)  ({\bf H3c}).
\end{itemize}
\end{quote}

The pairwise correlations of the dependent variable and the independent ones provide the correlation matrix (see Table \ref{tab:CorrM}).

\begin{table}[!h]
\centering
\begin{tabular}{l|rrrrrrrrr|}
                           & \multicolumn{1}{c}{\rotatebox{90}{$\log(\Delta t_C)$}} & \multicolumn{1}{c}{\rotatebox{90}{$\log(N_{A,e})$}} & \multicolumn{1}{c}{\rotatebox{90}{$\log(I_{\%A,e})$}} & \multicolumn{1}{c}{\rotatebox{90}{$\log(N_{T,e})$}} & \multicolumn{1}{c}{\rotatebox{90}{$\log(E_{\textrm{sim},e})$}} & \multicolumn{1}{c}{\rotatebox{90}{$\log(N_{O,e})$}} & \multicolumn{1}{c}{\rotatebox{90}{$\log(E_{C,e})$}} & \multicolumn{1}{c}{\rotatebox{90}{$\log(N_{G,e})$}} & \multicolumn{1}{c|}{\rotatebox{90}{$\log(N_{E_G,e})$}} \\ \hline
$\log(\Delta t_C)$         & \hspace{0.29cm}1.00                                    &                                                     &                                                       &                                                     &                                                                &                                                     &                                                     &                                                     &                                                        \\
$\log(N_{A,e})$            & \hspace{0.29cm}0.11                                    & \hspace{0.29cm}1.00                                 &                                                       &                                                     &                                                                &                                                     &                                                     &                                                     &                                                        \\
$\log(I_{\%A,e})$          & -0.07                                                  & -0.27                                               & \hspace{0.29cm}1.00                                   &                                                     &                                                                &                                                     &                                                     &                                                     &                                                        \\
$\log(N_{T,e})$            & \hspace{0.29cm}0.07                                    & \hspace{0.29cm}0.08                                 & -0.59                                                 & \hspace{0.29cm}1.00                                 &                                                                &                                                     &                                                     &                                                     &                                                        \\
$\log(E_{\textrm{sim},e})$ & \hspace{0.29cm}0.74                                    & \hspace{0.29cm}0.06                                 & \hspace{0.29cm}0.01                                   & \hspace{0.29cm}0.04                                 & \hspace{0.29cm}1.00                                            &                                                     &                                                     &                                                     &                                                        \\
$\log(N_{O,e})$            & -0.23                                                  & -0.03                                               & \hspace{0.29cm}0.05                                   & \hspace{0.29cm}0.01                                 & \hspace{0.29cm}0.02                                            & \hspace{0.29cm}1.00                                 &                                                     &                                                     &                                                        \\
$\log(E_{C,e})$            & -0.60                                                  & \hspace{0.29cm}0.023                                & -0.02                                                 & \hspace{0.29cm}0.01                                 & -0.53                                                          & \hspace{0.29cm}0.33                                 & \hspace{0.29cm}1.00                                 &                                                     &                                                        \\
$\log(N_{G,e})$            & -0.16                                                  & \hspace{0.29cm}0.01                                 & -0.07                                                 & -0.02                                               & -0.42                                                          & \hspace{0.29cm}0.19                                 & \hspace{0.29cm}0.23                                 & \hspace{0.29cm}1.00                                 &                                                        \\
$\log(N_{E_G,e})$          & -0.12                                                  & 0.00                                                & -0.07                                                 & \hspace{0.29cm}0.07                                 & -0.11                                                          & \hspace{0.29cm}0.42                                 & \hspace{0.29cm}0.43                                 & \hspace{0.29cm}0.14                                 & \hspace{0.29cm}1.00                                    \\ \hline
\end{tabular}
\caption{Correlation matrix of dependent and explanatory variables.}
\label{tab:CorrM}
\end{table}

With the explanatory variables of our model being defined, we are in position to formulate the econometric model by developing the equation $\eqref{eq:EcoModel}$:

\begin{align}
    \log(\Delta t_{C,e}) = \zeta &+ \kappa_{N_A} \cdot \log(N_{A,e}) + \kappa_{I_{\%A}} \cdot \log(I_{\%A, e}) + \kappa_{N_T} \cdot \log(N_{T,e}) \nonumber\\ 
    &+ \kappa_{E_\textrm{sim}} \cdot \log(E_{\textrm{sim},e}) + \kappa_{N_O} \cdot \log(N_{O,e}) + \kappa_{E_C} \cdot \log(E_{C,e}) \nonumber\\
    &+ \kappa_{N_G} \cdot \log(N_{G,e}) + \kappa_{N_{E_G}} \cdot \log(N_{E_G,e}) \nonumber\\
    &+ \varepsilon_e 
\label{eq:OLS1}
\end{align}\\

    Model validation is performed as follows. When handling a multivariate regression, one must pay particular attention to  multi-collinearity between the $Z_k$'s, which may distort the model. For that, the variance inflation factor (VIF) resulting from the regression of the explanatory variable $Z_k$ on the other explanatory variables which provide $R_{k}^2$, must be computed. The $\textrm{VIF}_k$ is then given as $\textrm{VIF}_k = 1/(1-{R_k}^2)$ and must be $< 10$ \cite{asteriou_applied_2015}. The stability of the variance has to be examined, namely by studying heteroskedasticity, which is ruled out if the $p$-value obtained from a White test is lower than a threshold $\alpha = 0.05$ \cite{asteriou_applied_2015}. The computation steps are performed with the \verb?Python? libraries \verb?statsmodels.api.OLS? for the regression, \verb?statsmodels.stats.outliers_influence? for the VIF and \verb?statsmodels.stats.diagnostic? for the White test.

%% file: results.tex
\section{Results}
\label{sec:results}
In order to establish evidence of collective action as an efficient way for tackling time-critical cybersecurity threats, we have resorted to data the MISP instance, which is run by the computer Incident Response Center Luxembourg (CIRCL). We used  a multivariate cross-sectional regression analysis of {\it completion time} (i.e., performance) required to characterize a threat event with both event related and collective action explanatory variables. 

\begin{table}[h!]
\centering
\begin{tabular}{lllr}
\hline
\multicolumn{2}{l|}{Dep. Variable}                 & \multicolumn{2}{l}{Completion Time}                                         \\ \hline
Method           & \multicolumn{1}{l|}{OLS}        & F-Stat             & \multicolumn{1}{l}{\hspace{0.29cm}$2.251 \times 10^3$} \\
No. Observations & \multicolumn{1}{l|}{22423}      & Prob (F-Stat)      & \multicolumn{1}{l}{\hspace{0.29cm}0.00}                \\
R-squared        & \multicolumn{1}{l|}{0.413}      & Log-likelihood     & \multicolumn{1}{l}{$-5.030 \times 10^4$}               \\ \hline
                 & \multicolumn{2}{l}{coeff}                            & \multicolumn{1}{l}{std err}                            \\ \hline
Const            & \multicolumn{2}{l}{$16.505^{(***)}$}  & $0.135$                                                \\
CountAttr        & \multicolumn{2}{l}{\hspace{0.29cm}$0.230^{(***)}$}   & $0.011$                                                \\
InhPer           & \multicolumn{2}{l}{${-0.089}^{(***)}$}               & $0.014$                                                \\
NTags            & \multicolumn{2}{l}{\hspace{0.29cm}${0.951}^{(***)}$} & $0.090$                                                \\
CumOrgs          & \multicolumn{2}{l}{${-0.346}^{(***)}$}               & $0.024$                                                \\
CumCompE         & \multicolumn{2}{l}{${-0.629}^{(***)}$}               & $0.006$                                                \\
NbGalaxies       & \multicolumn{2}{l}{${-0.083}^{(***)}$}               & $0.019$                                                \\
NbEventsinhisG  & \multicolumn{2}{l}{\hspace{0.29cm}${0.160}^{(***)}$} & $0.005$                                                \\ \hline
Skew             & -0.011                          & Durbin-Watson      & \multicolumn{1}{l}{1.302}                              \\
Kurtosis         & 2.833                           & Cond No.           & \multicolumn{1}{l}{76.4}                              \\ \hline
\end{tabular}
\caption{Results of the ordinary least squares (OLS) regression with the explained variable \texttt{CompletionT} and the explanatory variables: \texttt{CountAttr, InhPer, NTags, CumOrgs, CumCompE, NbGalaxies} and \texttt{NbEventsinhisG}, namely the number of attributes per event, the inherited percentage of attributes per event, the number of tags per event, the cumulative number of organizations at the creation of the event $e$, the number of already completed events by the organization at the creation of his new event $e$, the number of galaxies at the creation of the event $e$ and the number of events populating these galaxies at the creation of the event $e$. For each explanatory variable, the autoregressor coefficient (in the column \texttt{coeff}), as well as its standard deviation (in the column \texttt{std err}) are provided. The significance of the explanatory variables is given by the $p$-value and its threshold, i.e. $p-\textrm{value} < 0.1: (*), < 0.05: (**) \textrm{ or } < 0.01: (***)$ and the goodness-of-fit by the \texttt{R-squared}. The other added information are not necessary for the evaluation of the model.} 
\label{tab:OLS Stats}
\end{table}

The regression results are shown in Table \ref{tab:OLS Stats}. Overall, the regression model is robust and explains $41.3\%$ of the variance ($R^2 = 0.413$). Testing for hypothesis 1, the model shows that indeed event complexity measured by the number of attributes \texttt{CountAttr} and tags \texttt{NTags} influences performance negatively, i.e., event characterization completion time is increased. Hypothesis H1 is supported. Regarding how collective action improves performance (H2), the model shows that overall performance (i.e., completion time reduced) is positively associated with the number of organizations participating in MISP: Hypothesis H2a is supported. Hypothesis H2b could not be tested as a result of unexplained strong multi-collinearity between \texttt{CumOrgs} and \texttt{SimEvents}. Turning to Hypothesis 3 (i.e., knowledge integration increases performance), we find that more experienced organizations perform better in reducing event completion time. Hypothesis H3a is supported. We also find that the proportion of attributes that an event $e$ inherits from previous events, i.e., from the MISP CIRCL knowledge base, also positively influences performance. Hypothesis H3b is supported. Finally, testing for hypothesis H3c, i.e., modularity, we find mixed results. While the number of MISP Galaxies, measuring the number of modular sub-systems, influences positively performance, the number of events recorded in MISP Galxies, measuring to some extent the intensity of modularity, influences performance negatively. Hypothesis H3b is only partially supported.

 We have checked for multi-collinearity of the explanatory variables. We computed the variance inflation factor (VIF) for each explanatory variables, which happens to be all smaller than 10. This implies that there is no evidence of multi-collinearity between the selected explanatory variables (c.f., Table \ref{tab:VIF}). We also controlled for heteroskedasticity, i.e., a possible instability of the variance by performing a White statistics tests. We obtained $p\textrm{-value} < 10^{-2}$, which implies that there is no heteroskedasticity in our model. The post-analysis for the VIFs and the White statistics test completely validate the used model and its results.

\begin{table}[h!]
\centering
\begin{tabular}{lll}
Explanatory variables                                       & Notation  & VIF  \\ \hline
Number of attributes per event                              & $N_{A,e}$          & 5.15 \\
Inherited percentage of attributes per event $e$            & $I_{\%A,e}$        & 1.67 \\
Number of tags per event $e$                                & $N_{T,e}$          & 1.03 \\
Cumulated number of organizations at the creation of $e$    & $F_{\textrm{cum},e}$ & 6.73 \\
Cumulated number of completed events at the creation of $e$ & $E_{C,\textrm{cum},e}$       & 3.28 \\
Cumulated number of galaxies at the creation of $e$         & $N_{G,\textrm{cum},e}$      & 1.12 \\
Cumulated number of events in galaxies at creation of $e$   & $N_{E_G, \textrm{cum}, e}$  & 2.02\\
\hline
\end{tabular}
\caption{Computation of the variance inflation factor (VIF) for the explanatory variables of the econometric model. The values of the VIF allows to detect the presence of multi-collinearity between the considered variables. As all values $\textrm{VIF} < 10$, there is no evidence of multi-collinearity between the explanatory variables. These results validate the econometric model.}
\label{tab:VIF}
\end{table}

%% file: discussion.tex
\section{Discussion}
\label{sec:discussion}
Organizations are increasingly encouraged to cooperate and share information to overcome cybersecurity threats. Investigating how collective action unfolds and brings performance on information-sharing platforms is necessary as cybersecurity threats have become increasingly time-critical. In other words, not only collective action shall be used to characterize threat events, it also must be used to characterize threat events before attacks unravel \cite{wagner_cyber_2019}. Here, we have investigated collective action on MISP, a popular open source threat intelligence platform, from the perspective of the time required to fully characterize an event as an objective function to be optimized (i.e., completion time or performance). We found that performance is negatively associated with event complexity (Hypothesis 1) and positively associated with collective action (Hypothesis 2). Indeed, as the number of organizations taking part to information-sharing on the MISP instance studied increased, the time required to complete the characterization of events decreased. This result informs on positive returns on scale, which necessarily exist given the increased adoption of MISP as well as other information-sharing platforms. Nevertheless, the mechanisms at work generating these economies of scale have remained unclear. We considered the perspective of knowledge integration \cite{tononi_consciousness_1998} as the collective action process at work to generate the ``the whole is more than the sum of its parts'' \cite{sornette2014much}. With hypothesis 3, we tested and verified organizational learning, knowledge integration and modularity as positively associated with performance.\\

edWhile event completion time is associated with explanatory variables pertaining to event complexity, collective action, and knowledge integration, we could not establish causality. Although this is a significant limitation to our model, we have organized our multivariate cross-sectional regression in a way that minimizes the risks of uncovering spurious dependencies between the explained variable on the one hand and the explanatory variables on the other hand. And the fact that all our explanatory variables are significant (at the exception of \texttt{SimEvents}, the number of simultaneously open events on MISP CIRCL at the
creation, which had to be excluded from the model), shows that our proposed theory on {\it collective action for tackling time-critical tasks} is comprehensive and altogether robust. Yet, the regression analysis approach remains exploratory. Indeed, it does not provide reliable information on which precise collective action mechanisms generate positive returns on scale. Building and testing fine-grained causal models of critical cascades in collective action, inspired from e.g. \cite{sornette2014much,maillart_aristotle_2019,muric_collaboration_2019}, may surely help better understand the activity, learning, knowledge integration and modularization paths of contributing organizations, as well as how they handle time as a particularly scarce resource \cite{maillart_quantification_2011}. Indeed, when tackling large amounts of time-critical tasks, such as cybersecurity threats or incidents, contingencies necessarily appear \cite{kuypers_designing_2018}, which may affect coordination between contributors, and as a result performance, either in a transient way or by triggering long-term instability through cascades of disorganization. At the meso-scale, our model does not account for affinities between events, organizations and the combined commonalities of events and organizations. Indeed, as for number of collective action online platforms, modular {\it Galaxies} on MISP show that some sub-communities of organizations have specific goals when tackling cybersecurity threats. These specific interests deserve further scrutiny. For instance, are the organizations contributing to a given MISP galaxy active in the same industry? If not, why do they share interest in similar threats? Considering MISP (or other information-sharing platforms) from the perspective of threats, we may investigate kinship between threats, as they most often share attributes. Questioning and perhaps predicting how attributes are ``transmitted'' from one event to others is likely to be key to anticipate threats and guide organizations in their search of (respectively contributions to) threat information. It may even help decide what information should be shared and with whom.

Finally, our results show that completion time as an objective function in collective action concerned with time-critical tasks can be optimized. This opens further perspectives for computational social science research. One may envision to use machine learning in order to recommend personalized precision strategies that optimize the organization of collective action and knowledge integration. This may help make best use of time as an increasingly critically scarce resource, especially in face of a looming tsunami of cybersecurity threats.

%% file: conclusion.tex
\section{Conclusion}
\label{sec:conclusion}
Information-sharing in cyber-security has become an increasingly common collective action practice. Yet, its benefits have so far remained unclear. We have investigated MISP, a commonly used open source threat sharing platform, and we found how building a critical mass of contributing organizations and of knowledge to be integrated from past threats brings significant economies of scale. Through collective action, security researchers overcome the challenge of characterizing cybersecurity threats, which appear to be increasingly time-critical. We find that performance, defined as the time needed to fully characterize a threat event, is (i) negatively influenced its own complexity, (ii) positively influenced by collective action, and (iii) positively by learning, knowledge integration and modularity. Our results also inform more generally on how collective action can be organized online at scale and in a modular way to overcome a large number of time-critical tasks.

%% file: acknowledgments.tex
\section*{Acknowledgements}
The authors thank the WEIS'2021 and WEIS'2022 anonymous reviewers for their useful comments. The authors acknowledge support from the members of the Computer Incident Response Center Luxembourg (CIRCL) for making their data available and for their technical support. One of the authors (S. Gillard) acknowledges financial support from the Military Academy at ETH Zurich. Another author (D. Percia David) acknowledges financial support from the Cyber-Defence Campus (armasuisse Science and Technology).

%% file: appendices.tex
\appendix
\section*{Appendices}
\section{MISP: Description and Data Retrieval}
\label{Appendix:MISP_description}

\subsection{Detailed Description of MISP}

MISP is a partially de-centralized system of communities (e.g., NATO MISP, CIRCL MISP). interacting more or less together across MISP instances. A MISP instance consists in the installation of the MISP software and the community database in which community members share and collect data. Similarly to {\it GIT},\footnote{\url{https://git-scm.com/}} \textbf{organizations} work on their own instance and synchronize with remote instances. According to their sharing setting (i.e., your organization only, community only, connected communities, all communities or defined sharing group), community members have access to a certain amount of data.\\

Based on investigation needs or reports found in the newspapers or on specialized websites, the user creates an \textbf{event} to contextualize and encapsulate the related \textbf{attributes} (i.e., IoCs) and their properties (e.g., an IP address). All events have some general properties of the event, such {\it creation date}, aforementioned sharing level, {\it threat level} (i.e., 1: High, 2: Medium, 3: Low, 4: Undefined),  \textbf{analysis level} (i.e., 0: Initial, 1: Ongoing, 2: Complete) and a general description. The creator of an event can choose if this event is published on the remote instance or remains internal to the organization. Then, when the event is created, some attributes are added to populate this event. The event attributes refer to intrusion artifacts or methods used by  attackers. These attributes provide details and they are characterized by their \textbf{type} (e.g., filename|md5, sha256, etc.) and their belonging to a \textbf{category} (e.g., Antivirus detection, Targeting data, etc.), putting them in the context and justify then its attribution to its corresponding event. To add an attribute related to an event, global information such as its category, its type and its distribution, either the same as for the event or its own rule, is required, as well as two important text fields: \textbf{value} and \textbf{contextual comment}. The "value" field stores the data we want to add, e.g. an url leading to a report, while the ``comment'' field allows complementary information about the attribute. Moreover, it is possible to allocate one \textbf{tag} or more to an event in order to simplify the read and the classification of this event. These tags can follow the MISP taxonomy, i.e. a fixed machine-tag vocabulary, or be created by the users according to their needs.\\

On the platform, events, attributes, organizations and tags are associated to their own identification (ID) number and their creation are timestamped, as well as the publication and the last update of an event.\\

As an open-source platform, MISP relies on voluntary action. On the one hand, its members can create or exchange content. On the other hand, these same actors can obtain new insights or possible response elements from the community regarding cyber-threats of interest. To organize interactions and to create information-sharing incentives for the participants, MISP offers several aforementioned sharing levels through a comprehensive sharing model. Users can select to whom they want to share information among the following levels from the most restrictive to the most open. Regardless of access and to guarantee the quality of the shared data, only organizations that created an event have the permission to modify this event. However, each user has the possibility to submit his own suggestions to change an event created by others, who can then accept or reject the proposal.\\

Moreover, the experience of older MISP versions has shown that the time to fill the fields and a complicated web interface introduce some frictions. For this purpose, a free text importer has been deployed, so that data can be copied and pasted into the intended field. Further, MISP implements a heuristics-based algorithm, which helps users to match events or event attributes with events or attributes from events already in the data base. However, let us added that the matching is never performed automatically, and goes through human supervision.\\

   \subsection{Data Retrieval}
    To investigate our hypotheses, we have to curate the main dataset by considering only the closed events, i.e. the events with an analysis level equal to $2$, meaning ``complete''.

    To retrieve the data, we have followed the user guide\footnote{\url{https://www.circl.lu/doc/misp/book.pdf}} provided by the MISP CIRCL instance. We used the PyMISP module to download data in \texttt{.json} format file. The main dataset contains one file per event. These event files contain the attributes (see MISP core format\footnote{\url{https://www.misp-standard.org/rfc/misp-standard-core.html}}), as well as the name and the ID of the concerned organizations. However, due to the policy of the MISP CIRCL instance, we cannot disclose the names of these organizations and present no interest and have no influence on the obtained results.\\

\section{Exploratory Analysis of the Data Set}
\label{sec:ExploA}

    \subsection{Probabilistic Distributions}
    \label{subsec:Probadist}
    In order to understand the mechanisms handling on the MISP platform, we want to investigate the distribution of our data, we have to present the selected variables and explore the distribution associated with these. In some cases, we are able to investigate the probabilities distribution. Hence, if we consider a random variable $X$ with a probability density function (PDF) $f_X(x)$, the cumulative distribution function (CDF), $F_X(x)$ is given by:
     
    \begin{align}
    F_X(x) = P(X \leq x) = \int_{-\infty}^x f_X(t) dt.
    \label{eq:CDF}
    \end{align}
     
     Then, thanks to the formula $\eqref{eq:CDF}$, the complementary cumulative distribution function (CCDF) $\overline{F_X}(x)$ can be written as follow:
     
    \begin{align}
    \overline{F_X}(x) = 1 - F_X(x) = P(X > x).
    \end{align}
    \noindent
    This CCDF provides a rank ordering of the selected variables.\\

    \subsection{Fit of the Data}
    
    Before we start fitting our data, a visual analysis can be performed. Then, in any case, by varying the scale of axis -- double linear, linear-logarithmic or double logarithmic -- depicting our data, we are able, if our data follow approximately a straight line in one of cases presented below, to fit the data. The logarithmic scales are considered in base 10.\\
     
    \subsubsection{Double Linear Scales}
    \label{subsub:lin-lin}
    
    By considering two vectors of data $\overrightarrow{x}$ and $\overrightarrow{y}$ and plotting the data contained in $\overrightarrow{y}$ ($y$-axis) in function of the data in $\overrightarrow{x}$ ($x$-axis) in linear scale for the axes $x$ and $y$. If the displayed data shows an approximate straight line, that means that each element $y_i$ of the vector $\overrightarrow{y}$ is given by the relation:
    
    \begin{align}
    y_i = a \cdot x_i + b,
    \label{eq:lin-lin}
    \end{align}
    
    where $a$ is the slope of the straight line and $b$, its intercept. Thanks to the relation $\eqref{eq:lin-lin}$, we are able to compute the estimated $\hat{y_i}$, $a$ and $b$ by applying a least-square linear regression. To validate the parameter obtained from the linear regression, we need to establish the goodness-of-fit with these parameters. For this type of simple linear regression, we use the Pearson's coefficient of determination $R^2$ and, to reinforce the results of $R^2$, we perform a Wald test with a chosen level $\alpha = 0.05$ to define if these two samples are significantly identical or not. Then a value $|R^2| ~\approx 1$ implies a strong correlation between $\overrightarrow{x}$ and $\overrightarrow{y}$, while a $p\textrm{-value} < \alpha$  for the Wald test allows us to affirm that the parameters of the fit are good and the estimated $\hat{\overrightarrow{y}}$ are significant according to $\overrightarrow{y}$. With these indicators, we can thus say that our data have a linear behaviour which follow a straight line with slope $a$. $a$ is the most important parameter for our analysis, then $b$ can be neglected 
    To produce the linear regression on our data and to compute $R^2$ and the $p\textrm{-value} < 0.05$ for the Wald test, we use the \verb?Python? library \verb?scipy.stats.linregress?.\\
    
    \subsubsection{Linear-Logarithmic Scales}
    \label{subsub:lin-log}
    
    Following the same process as above, excepted that we put the $y$-axis in logarithmic scale. If data $\overrightarrow{y}$ in function of $\overrightarrow{x}$ depict a straight line, we can write the relation as: 
    
    \begin{align}
    \log(y_i) &= a \cdot x_i + b, \textrm{derived from} \label{eq:lin-log}\\
    y_i &= 10^{(a \cdot x )} \cdot 10^b \label{eq:exp},
    \end{align}
    
    where, $a$ is the slope or the increasing factor and $b$ the intercept or an additive constant depending on the relations $\eqref{eq:lin-log}$ and $\eqref{eq:exp}$. In this case, the data describe an exponential shape. As this process is not used in this article, we don't develop completely this, it remains nevertheless important to pursue with the last case.\\
    
    \subsubsection{Double Logarithmic Scales}
    Considering the same method than the two aforementioned cases, we plot the data contained in $\overrightarrow{y}$ versus $\overrightarrow{x}$ on logarithmic $x$- and $y$-axis. In the case where the data behave itselves like a straight line we are then able to deduce the relation:
    
    \begin{align}
    \log(y) &= a \cdot \log(x) + b, \textrm{derived from} \label{eq:log-log}\\
    y &= x^a \cdot 10^b \label{eq:pl},
    \end{align}
    
    where $a$ is the slope or the exponent and $b$ is the intercept or a multiplicative constant according to the equations $\eqref{eq:log-log}$ and $\eqref{eq:pl}$. From the relation $\eqref{eq:log-log}$, we can determine the estimated values for elements $\hat{y_i}$, $a$ and $b$.\\
    
    From here, we have to distinguish the two following cases:
    
    \begin{align}
    \begin{cases}
    a \geq 0 \textrm{ or}\\
    a < 0
    \end{cases}
    \end{align}
    
    In the case of $a \geq 0$, we treat a power function given by the equation $\eqref{eq:pl}$. The fit can be, as for the double linear case, obtained by performing the least-square linear regression. Then, the goodness-of-fit is given by the Pearson's coefficient of determination $R^2$ and the $p\textrm{-value} < 0.05$ for the Wald test. The results are computed the \verb?Python? library \verb?scipy.stats.linregress?. 
    
    In the case of $a<0$, we are in presence of a power law. Due to the presence of the logarithm on both sides of $\eqref{eq:log-log}$, we cannot apply a least-square linear regression, because this method and the similar ones return systematic errors for common conditions. For this reason, it is impossible to trust the results \cite{clauset_power-law_2009}. Instead of this method, we estimate the parameters $a$ with the method of maximum likelihood after a quadratic approximation to the log-likelihood to deal with our discrete values. In our analysis, the parameter $b$ is not relevant and we don't need to estimate this. To determine if it really handles of a power law, we proceed to a Kolmogorov-Smirnov test, attempting to minimize the distance between the estimated parameters and our data. If the $p$-value from the Kolgomorov-Smirnov is smaller than the chosen threshold $\alpha = 0.05$ , we can affirm that our data follow a power law \cite{clauset_power-law_2009}.
    Sometimes, the fits don't fit very well with a power law distribution that is why we have to investigate other heavy-tailed distributions like the log-normal (L) or the Weibull (W) (i.e., stretched-exponential) distributions, for which we can define the goodness-of-fit with the previous Kolmogorov-Smirnov test and its $p$-value. However, with approximately same results, the power law is privileged because it is determined by one parameter instead of two parameters for the two aforementioned distributions.\\
    The computations in this part have been widely inspired from the works of A. Clauset \& al.  and done with \verb?Python? libraries such that \verb?plfit? for the powerlaw and implemented according to the works of A. Clauset \& al. for the other distributions \cite{clauset_power-law_2009}.\\
   
\subsubsection{Goodness-of-fits Summary}
\label{subsub:GofS}

The results for the fits presented in this article (i.e., Figure \ref{fig:org}, \ref{fig:CCDFs} and \ref{fig:Deltat}), as well as their goodness of are detailed in the below Table \ref{tab:Gofs}.

\begin{table}[!h]
\centering
\begin{tabular}{llllll}
\hline
Fig & Model    & Estimated Parameter(s)   & Goodness-of-fit       & $p$-value   & Quality \\ \hline
1A  & PL  $^a$ & $\mu_{\mathrm{att}} = 0.64(1)$   & $6.43 \times 10^{-2}$ & $< 10^{-2}$ & (***)                                  \\[5pt] \hline
1B  & PL $^a$  & $\mu_{\mathrm{tags}} = 2.26(6)$  & $1.52 \times 10^{-1}$ & $<10^{-2}$  & (***)                                  \\[5pt] \hline
2A  & PL $^a$  & $\mu_{\mathrm{events}} = 0.54(4)$  & $1.51 \times 10^{-1}$ & $<10^{-2}$  & (***)                                  \\[5pt] \hline
2B  & LR $^b$  & $\beta_{\mathrm{orgs}} = 0.79(1)$                                    & 0.99                  & $<10^{-2}$  & (***)                                  \\[5pt] \hline
3A  & LR $^b$  & $\beta_{\Delta} = -0.93(1)$                                         & 0.97                  & $<10^{-2}$  & (***)                                  \\[5pt] \hline
3B  & LR $^b$  & $\beta_{\Delta}^1 = (-6.32 \pm 0.91) \times 10^{-2}$                  & 0.86                  & $<10^{-3}$  & (***)                                  \\[5pt]
    &          & $\beta_{\Delta}^2 = (-7.12 \pm 0.59) \times 10^{-2}$ &                     &                       &                      \\ \hline
\end{tabular}
\caption{Goodness-of-fits summary. The fits are generated by the Power Law $^a$  and ordinary least squares (OLS) Linear Regression $^b$ models. The goodness-of-fit are obtained with the Pearson's coefficient $R^2$ $^a$ and the $p$-value of a Wald test for the Linear Regression $^a$ model and with the Kolmogorov-Smirnov statistic test, also providing the $p$-value, for the Power Law $^b$ model. The results are computed with the \texttt{Python} libraries \texttt{scipy.stats.linregress}$^a$ and \texttt{plfit} $^b$.}
\label{tab:Gofs}
\end{table}